\documentclass[pdflatex,sn-mathphys-num]{sn-jnl}

\usepackage{graphicx} \usepackage{comment}
\usepackage{siunitx}
\usepackage{hyperref}

\usepackage{amsmath, amsfonts, amssymb}
\usepackage{upgreek}
\usepackage{bm}

\newcommand{\D}{\mathrm{d}}
\newcommand{\Eb}{\mathbf{E}}
\newcommand{\Bb}{\mathbf{B}}
\newcommand{\vb}{\mathbf{v}}
\newcommand{\ssb}{\mathbf{s}}
\newcommand{\Omegab}{\bm{\Omega}}

\newcommand{\m}{\unit{\metre}}
\newcommand{\cm}{\unit{\centi\metre}}
\newcommand{\mm}{\unit{\milli\metre}}
\newcommand{\um}{\unit{\micro\metre}}
\newcommand{\nm}{\unit{\nano\metre}}
\newcommand{\ps}{\unit{\pico\second}}
\newcommand{\fs}{\unit{\femto\second}}

\newcommand{\Hz}{\unit{\hertz}}
\newcommand{\MHz}{\unit{\mega\hertz}}
\newcommand{\mins}{\unit{\minute}}
\newcommand{\hrs}{\unit{\hour}}
\newcommand{\ccm}{\unit{\centi\metre^{-3}}}
\newcommand{\mA}{\unit{\milli\ampere}}
\newcommand{\Amp}{\unit{\ampere}}
\newcommand{\nC}{\unit{\nano\coulomb}}
\newcommand{\pC}{\unit{\pico\coulomb}}

\newcommand{\MeV}{\unit{\mega\electronvolt}}
\newcommand{\GeV}{\unit{\giga\electronvolt}}
\newcommand{\MVm}{\unit{\mega\volt/\metre}}

\newcommand{\Tm}{\unit{\tesla\metre}}

\begin{document}

\title[An electron injector for the EIC based on proton-driven plasma wakefield acceleration]{An electron injector for the Electron-Ion Collider\\based on proton-driven plasma wakefield acceleration}

\author*[1]{J.~P.~Farmer}\email{j.farmer@cern.ch}
\author*[2]{H.~Jaworska}\email{helena.jaworska@postgrad.manchester.ac.uk}
\author[1]{A.~Caldwell}
\author[3]{N.~Lopes}\author[2]{A.~Pukhov}\author[2,4]{L.~Reichwein}\author[5]{F.~Willeke}\author[6,7]{M.~Wing}

\affil[1]{\orgname{Max Planck Institute for Physics}, \orgaddress{\city{Garching}, \country{Germany}}}
\affil[2]{\orgname{Heinrich-Heine-Universit\"at}, \orgaddress{\city{D\"usseldorf}, \country{Germany}}}
\affil[3]{\orgname{Instituto Superior Tecnico}, \orgaddress{\city{Lisbon}, \country{Portugal}}}
\affil[4]{\orgname{Forschungszentrum J\"ulich}, \orgaddress{\city{J\"ulich}, \country{Germany}}}
\affil[5]{\orgname{Brookhaven National Laboratory}, \orgaddress{\city{Brookhaven}, \country{US}}}
\affil[6]{\orgname{UCL}, \orgaddress{\city{London}, \country{UK}}}
\affil[7]{\orgname{DESY}, \orgaddress{\city{Hamburg}, \country{Germany}}}

\abstract{
We describe an electron bunch injector scheme based on proton-driven plasma wakefield acceleration for the Electron-Ion Collider.  The proton bunches are assumed to have the same energy as those used in the EIC, and the polarized electron source is that used in the current EIC design.  We describe the different elements making up the injection scheme and give an estimate for the performance. Using a novel top-up injection scheme for electrons, the charge in the ring is accumulated over many injection events, avoiding deleterious beamloading in the proton-driven wakefield stage and allowing the polarization in the ring to be maintained. Our initial study indicates that for electron bunches at 10~\GeV, 1160 bunches can be stored in the ring over a 10-hour period with an average bunch charge of 19.3~\nC\ and an average polarization of 75.8~\%, representing 69~\% of the EIC design specification for charge and exceeding the specification for polarization.  Operation at 18~\GeV\ is achieved by shortening the electron beam lifetime to increase the average polarization.  Paths to improve upon these results are identified.
}


\maketitle

\section{Introduction}
\begin{figure}[hbpt]
    \centering
    \includegraphics[width=0.5\linewidth]{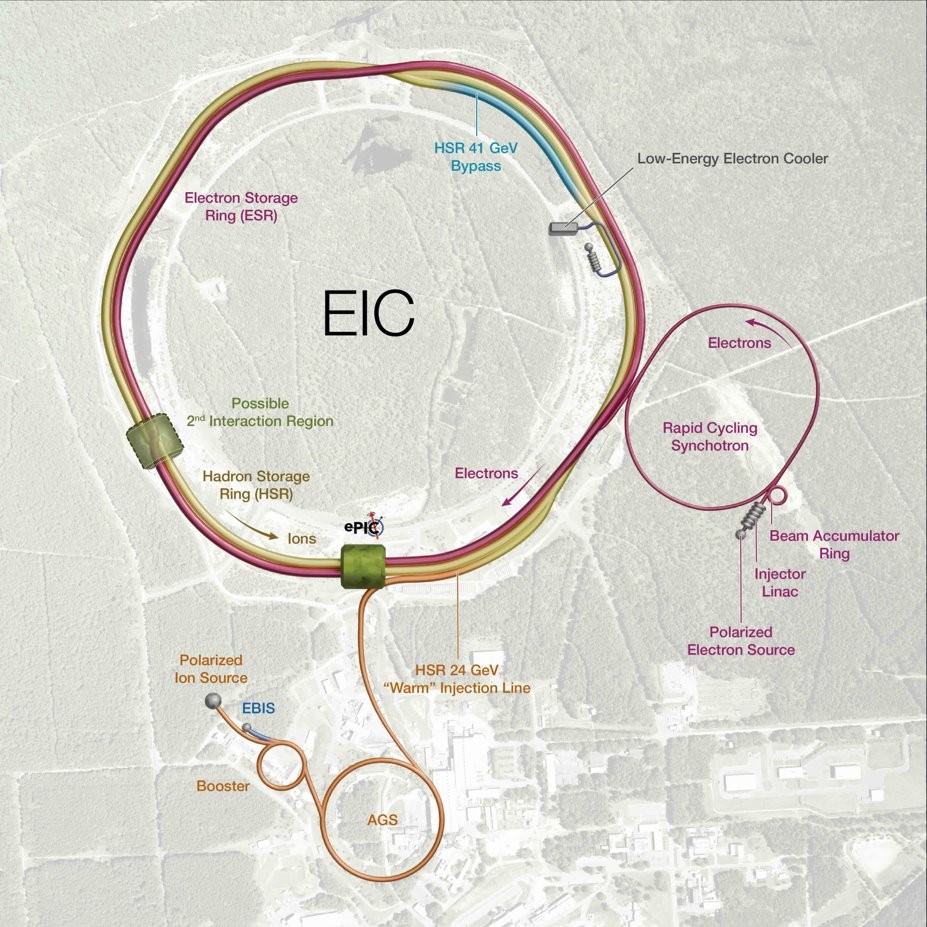}
    \caption{The planned layout for the  EIC accelerator complex.}
    \label{fig:RHIC}
\end{figure}

The Electron-Ion Collider (EIC) is planned to be constructed at Brookhaven National Laboratory (BNL)\cite{accel-willeke-eic_cdr}. The facility will make use of the existing infrastructure of the Relativistic Heavy Ion Collider (RHIC), in particular one of its two superconducting accelerator rings (see the schematic layout in Fig.~\ref{fig:RHIC}). The EIC is designed to deepen our understanding of nuclear structure, provide insight into the phenomenon of increasing gluon density at low Bjorken-$x$ -- also referred to as gluon saturation -- and help resolve the long-standing question of the proton spin.

This ambitious scientific program requires a challenging accelerator design capable of supporting collisions between electrons with energies of up to 
$18$\,GeV and a wide range of ions, from protons to uranium, with energies of up to 
$275$\,GeV. In addition, the collider must deliver a high luminosity of up to 
$ L = 10^{34}$~cm$^{-2}$~s$^{-1}$, provide polarized electron and light-ion beams with an average polarization of $70\, \%$, support polarization reversal, and operate over a centre-of-mass energy range from $29$\,GeV to $140$\,GeV.
Furthermore, the design must accommodate a multipurpose colliding-beam detector with particular emphasis on the detection of forward- and backward-scattered particles emerging from the interaction point.

These demanding requirements must be met at an affordable cost, which constitutes a significant challenge in itself. This motivates the exploration of novel technical solutions capable of delivering high performance at moderate cost. In this work, we report on a concept for a full-energy injector capable of providing electron beams up to $18$\,GeV. The concept makes use of proton bunches, similar to those used in the EIC, to drive a compact and comparatively inexpensive plasma wakefield accelerator.  Such a system could accelerate electrons from energies of about $100$~\unit{\mega\electronvolt} to $18$~\unit{\giga\electronvolt}.

The study connects proton-driven plasma wakefield acceleration to a concrete collider application, and can inform future studies for the potential applications of this technology. 
It should be noted that this study originally envisaged retaining the RHIC Blue Ring to accelerate proton bunches.  However, at time of publication, the Blue Ring is being dismantled as part of the ongoing construction of the EIC.  The source of the proton beam is therefore uncertain.

\subsection{EIC baseline design}\label{sec:EIC}
We begin with an overview of the EIC baseline specification, against which the proposed scheme must be compared.  The EIC baseline includes very high electron beam currents of up to 
\SI{2.5}{\ampere} and a high beam polarization, exceeding the equilibrium polarization achievable in a storage ring. Electrons are produced with a polarization of \SI{85}{\percent} by the existing polarized gun. Their polarization is preserved during acceleration to the final energy before injection into the Electron Storage Ring (ESR), where it decays toward the equilibrium polarization of $50\,\%$. Stored electron bunches are planned to be replaced with fresh bunches once their polarization drops below approximately \SI{60}{\percent}. This quasi-continuous electron injection (corresponding to a rate of about \SI{1}{\hertz} for the \SI{18}{\giga\electronvolt} case) places strong demands on the capabilities of the electron injector. The corresponding electron beam parameters are summarized in Table~\ref{tab:ESR-params}.

\begin{table}[hbpt]
\centering
\begin{tabular}{|l|c|c|c|}

\hline  
Parameter            & Units      & \multicolumn{2}{|c|}{Value} \\ \hline 
Electron energy      & \GeV       & 10                     & 18                     \\
\hline
Maximum bunch charge & \nC        & 28                     & 10.7                   \\
Number of bunches    &            & $1160$                 & $290$                  \\
Beam Current         & \Amp       & 2.5                    & 0.24                   \\ 
Initial polarization &  \%        & 85                     & 85                     \\
Average polarization &  \%        & 70                     & 70                     \\
Horizontal emittance &  \nm       & 20                     & 24                     \\
Vertical emittance   &  \nm       & 1.3                    & 2.0                    \\
Energy loss per turn &  \MeV      & 4.01                   & 42                     \\
Relative RMS energy spread    & $10^{-4}$ & 5.5                    & 10                     \\ 
Polarization lifetime& \mins      & 406                    & 20.4                   \\ 
\hline 
\end{tabular}
\caption{Design parameters of the EIC electron beam for both 10 and 18~\GeV\ ESR operation.}
\label{tab:ESR-params}
\end{table} 
\subsection{Overview of the proposed scheme}\label{sec:overview}
We consider a scheme based on proton-driven plasma wakefield acceleration as the basis for an electron injector for the EIC. Since currently available proton bunches are long compared to the required plasma period, we make use of plasma to pre-condition the proton driver, a method developed by the Advanced Wakefield Experiment (AWAKE) Collaboration~\cite{pwfa-AWAKE-symmetry}, discussed in Section~\ref{sec:AWAKE}. We note that many of the challenges presented in the paper for the plasma-based acceleration part have been solved for the AWAKE case. This is a first attempt to develop a proton-driven plasma acceleration scheme for the electron bunches for the EIC, which can then be optimized and worked out in more detail in the future should interest develop in pursuing this approach.

There are some differences between the proposed scheme and AWAKE, which we discuss here to set the stage for the following more detailed sections. In our approach, the proton bunches which drive the plasma wake are delivered by a second proton ring with parameters as given in Table~\ref{tab:protons}.  A more detailed description of the proton bunch preparation scheme is presented in Section~\ref{sec:proton}.

\begin{table}[hbpt]
    \centering
    \begin{tabular}{|c|c|c|}
    \hline
       Parameter                 & Units  & Value   \\
    \hline
        Energy                       & \GeV & 275  \\
        Bunch population             &      & \num{3e11}  \\
        Number of bunches            &      & 1160  \\
        Bunch length                 & \mm & 50 \\
        Bunch waist  & \um & 200\\
Normalised transverse emittance & \um & 2 \\
        Ring cycling time            & \mins & 11 \\
    \hline
    \end{tabular}
    \caption{The parameters of the proton bunches from the second proton ring taken in our studies.}
    \label{tab:protons}
\end{table}

The electron bunches are prepared using the existing polarized electron source~\cite{spin-litvinenko-conventional}.  Since there is an anti-correlation between the electron bunch charge and the maximum energy to which the electrons can be accelerated in the plasma, we have chosen $E_e=10$~\unit{\giga\electronvolt} as the working point for our studies, as this is expected to be the most difficult set of parameters to achieve given the requirement of high electron bunch charge. These considerations lead to the electron bunch parameters given in Table~\ref{tab:electrons}.

\begin{table}[hbpt]
    \centering
    \begin{tabular}{|c|c|c|}
\hline
       Parameter  & Units  & Value   \\
\hline
    Energy entering plasma   & \MeV & $150$  \\
    Energy after plasma     & \GeV & $10$  \\
Bunch charge            & \unit{\nano\coulomb} & 1 \\
    Bunch length & \um & $90$\\
    Normalised transverse emittance & \um & 10\\
    Bunch waist at plasma entrance & \um & 12.8\\
\hline

    \end{tabular}
    \caption{The parameters of the electrons bunches taken in our studies.}
    \label{tab:electrons}
\end{table}

Note that the electron bunch charge for one acceleration event is approximately 28 times lower than the value required for EIC physics (see Table~\ref{tab:ESR-params}).  To achieve the required charge of $28$~\nC, we foresee using `top-up' injection, i.e., a fresh bunch of $1$~\nC\ would be added to already circulating bunches at an average rate of $\sim 1$~\Hz.  This can be accomplished by extracting proton bunches in bursts at, e.g., $20$~\Hz\ from the proton ring for electron bunch acceleration.  Given the proton ring cycling time, this allows for an average rate of $1.75$~\Hz\ of full-energy electron bunches for injection into the ESR. Note that this would not only keep the needed bunch charge at the required level, but would also allow for maintaining a high average polarization of the electron bunch.  The electron pre-injector, based on conventional RF technology, is described in Section~\ref{sec:electron}, and the top-up injection scheme in Section~\ref{sec:topup}.

As with AWAKE, two plasma sections are necessary: the first to correctly condition the drive beam, and a second for acceleration.
A laser-ionized alkali source will be used for the first plasma, allowing the controlled self-modulation of the proton beam into a train of microbunches.  A discharge plasma source is used for the second plasma, allowing the required plasma length to reach 10 GeV.  A description of the plasma sources is given in Section~\ref{sec:PlasmaSource}.

The modulation of the proton beam in the first plasma and the acceleration of the electron bunch in the second plasma have been studied using particle-in-cell simulations as described in Section~\ref{sec:PlasmaAccStudies}.  The studies are based on extensive developments performed in the context of the AWAKE experiment and are based on codes which have been verified to be accurate for this scheme of plasma wakefield acceleration~\cite{pwfa_farmer_tearing}.  We note that the electron bunch polarization is expected to be maintained during the acceleration process, as discussed
in Section~\ref{sec:PlasmaPolarized}.   The electron bunches exiting the plasma must be finely tuned in energy to match the set energy of the ESR.  This fine-tuning, as well as the top-up injection scheme into the ESR are described in Section~\ref{sec:TuneandInject}.

We discuss the layout of the main components that would be necessary for this electron injector scheme in Section \ref{sec:PotentialLayout}, and discuss possible locations for installation in the EIC tunnel.  We then conclude with a summary of the results found in our studies to date, and discuss ways in which the results could be enhanced.
 
\subsection{The AWAKE Experiment}\label{sec:AWAKE}
Plasma can sustain high electric fields and can be used to produce accelerating gradients larger than in conventional particle accelerators~\cite{PhysRevLett.43.267,PhysRevLett.54.693,Blumenfeld:2007ph,Gonsalves:2019wnc,Albert_2021}.  First experiments relied on a laser pulse or an electron bunch to disturb the plasma and drive large fields in their wake.  Proton drivers were proposed~\cite{Caldwell:2008oob} as a means to reach high, TeV-level, energies in one plasma stage.  These initial simulations relied on high-energy protons in short bunches of order 100~\unit{\micro\metre}.  Although the creation of short proton drive bunches is being actively pursued~\cite{pwfa-farmer-higgs,accelerator-willeke-FFA}, they are not currently available at any laboratory.

High-energy bunches of protons with lengths of $\mathcal{O}(\rm cm)$ such as those at CERN, BNL and FNAL are available, although such bunches are expected to drive only weak wakefields.  To overcome this, the self-modulation instability~\cite{Kumar:2010bc} can be utilised to divide a long driver bunch into a group of short so-called microbunches  that can resonantly drive wakefields.  The AWAKE~\cite{AWAKE:2014wjf,Caldwell:2015rkk} experiment located at CERN  makes use of this technique to investigate proton-driven plasma wakefield acceleration. 

AWAKE uses protons from the CERN Super Proton Synchrotron (SPS) with energy of \SI{400}{\giga\electronvolt} 
in  bunches of \mbox{$3 \times 10^{11}$\,protons} and $5-10$\,cm in length.  Microbunches are formed in a controlled manner by using the Relativistic Ionization Front (RIF) of a high-power laser to seed the self-modulation mechanism~\cite{pwfa-batsch-seeded}; alternatively, electron bunches can also be used to seed the self-modulation~\cite{AWAKE:2022kmf}.  The strong wakefields driven by these microbunches are sampled by a witness bunch of electrons that is externally injected and accelerated to higher energies.  The whole process currently takes place in a ten-metre-long rubidium plasma source of density in the range $10^{14} - 10^{15}$\,cm$^{-3}$.  The plasma is created by ionizing rubidium vapour at a temperature of about 200\,$^{\rm o}$C. Initial experiments have shown a clear microbunching of the long proton bunch~\cite{pwfa-AWAKE-acceleration,pwfa-batsch-seeded} and acceleration of electrons from 19\,MeV up to 2\,GeV within the 10\,m\ of plasma~\cite{pwfa-AWAKE-acceleration}.

\begin{figure}
    \centering
    \includegraphics[width=\linewidth]{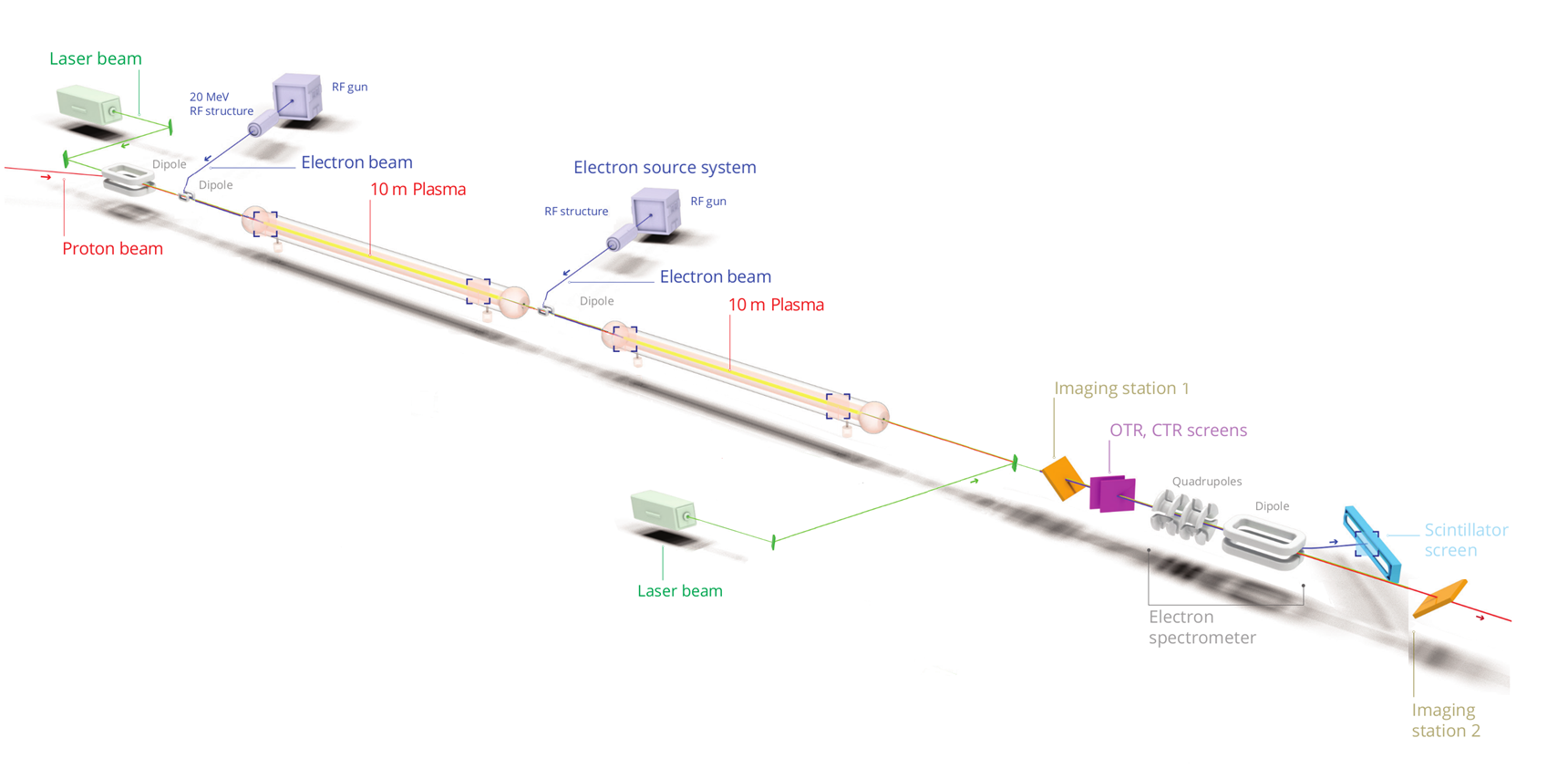}
    \caption{Overview of the layout of future AWAKE experiments showing two plasma sources~\cite{pwfa-AWAKE-symmetry}.  In the first  rubidium plasma source proton bunches can have the modulation seeded by a laser pulse or electron bunch.  The second plasma source is shown here as laser-ionized rubidium but could also be a discharge plasma source.  Various diagnostics are shown after the second plasma source.}
    \label{fig:AWAKE_layout}
\end{figure}

Future experiments at AWAKE~\cite{pwfa-AWAKE-symmetry} will focus on electron acceleration that is scalable to longer distances and with control of the electron bunch emittance.  After these experiments, scheduled to start in 2029, the AWAKE scheme will have been sufficiently demonstrated for application, such as that discussed here.  The AWAKE experiment will therefore be upgraded to have two plasma sources (see Fig.~\ref{fig:AWAKE_layout}), where the self-modulation of the proton bunch takes place in the first and electron acceleration in the second.  Rubidium vapour will again be used for the first plasma source at a default density of $7 \times 10^{14}$\,cm$^{-3}$, but with the ability to introduce a step in the density that will stabilize the wakefields~\cite{Lotov:2011ypq,pwfa-lotov-step}.  Electron bunches will be injected on-axis in a 30-cm gap between the first and second plasma source.  Electrons of 150 MeV in bunches of a few hundred pC in charge with a normalised transverse emittance of about \SI{2}{\micro\metre} and a duration \SI{200}{\femto\second} will be used.  These much shorter electron bunches than currently used ($>4$\,ps) should lead to a significantly higher capture of electrons in the wakefields and a more controlled acceleration process.  The second plasma source will be at least 10\,m in length and could be based on laser-ionized rubidium vapour or a discharge plasma source using a noble gas, where a discharge source should in principle be scalable to lengths longer than is possible for a rubidium vapour source.

The knowledge that has been and will be gained from the AWAKE experiment feeds directly into the design concept we envisage for the EIC electron injector.

\section{Conventional Accelerators}\label{sec:conventional}

In the next sections, we describe the RF-based accelerator components required to provide the driving proton beam for the plasma accelerator and the electron bunch to be accelerated.  These are the proton bunch accelerator to drive the plasma wake and the polarized electron bunch system needed to prepare the electron for acceleration in plasma.

\subsection{The proton accelerator}\label{sec:proton}

The RHIC Blue-Ring was a $250$~\GeV\ superconducting accelerator in the 3890~\m\ tunnel, one of the two RHIC accelerator rings.  The Blue-Ring superconducting magnets are configured for clockwise particle motion, the direction of the EIC electron beam.  As the Blue Ring was considered not to be needed for the EIC, this study considered retaining the ring into the EIC, with upgrades to the RF and power supply systems similar to those planned for the RHIC Yellow Ring.

We therefore assume a second 275~\GeV\ proton ring, similar to the EIC HSR, configured for beam delivery in the same direction as the EIC electron beam.  This ring provides proton bunches which generate the plasma wakefield which accelerate an electron beam from $150$~\MeV\ to up to 18~\GeV. 

As stated in Table \ref{tab:protons}, the assumed proton drive beam consists of $N_b = 1160$ bunches of $N_p = \num{3e11}$ protons at $275$~\GeV\ every $T_{cycle} = 11$~\mins.
This corresponds to a bunch current of $I_b = 3.7$~\mA\ and a total beam current of $I_{total} =4.3$~\Amp.
These performance requirements considerably exceed the beam parameters planned for the $275 $~\GeV\ EIC Hadron Storage Ring (HSR), for which the bunch population is $N_p = \num{6.3e10}$, corresponding to $I_b=0.78$~\mA, $I_{total} = 0.9$~\Amp.  The bunch length of 50~\mm\ and number of bunches for the proposed scheme is the same as the HSR, and the invariant longitudinal emittance is a factor two higher.

The EIC HSR proton beam is polarized and originates from an intensity-limited polarized proton source. The drive beam for plasma wakefield acceleration does not need to be polarized, and so protons would be generated in an unpolarized H$^{-}$ source that can provide four times the bunch intensity, enabling the required driver proton bunch population.  Such beam current will limit the $275$~\GeV\ storage time to less than a minute because of image current heating of the stainless steel beam pipe. The required longitudinal emittance can be achieved because the transverse emittance of the round driver beam can be considerably relaxed compared to the flat collider proton beam, 
$(\varepsilon_x\cdot\varepsilon_y)_{driver} = 10\times 
(\varepsilon_x\cdot\varepsilon_y)_{collider}$.

Using the H$^{-}$ source, the Alternating Gradient Synchrotron (AGS) will accelerate unpolarized bunches of \num{3e11} protons at a rate of 10~\Hz, corresponding to a total injection time of \SI{2}{\minute}. Bunches will be extracted from the proton ring at a rate of \SI{20}{\hertz}, corresponding to \SI{1}{\minute}. The ramp time is limited by the time constants of the quench protection circuitry. With a modest improvement in the electronics, a ramp cycle time of \SI{8}{\minute} can be achieved. Thus, the assumed driver cycling time of 11~\mins\ is possible.

During RHIC operation with polarized protons, with several hours between injections, the accelerator complex was sufficiently flexible to provide proton beams to other applications.  The same flexibility is expected for the EIC.  Although the shorter cycling time of the second proton ring will lead to increased scheduling competition, the fill time remains a small fraction of the total cycle and so the AGS is expected to remain available for other applications.

We conclude that a proton ring similar to the EIC HSR with some improvements to electronics and cycle time would be able to provide the drive beam for plasma wakefields that accelerate 1~\nC\ electron bunches for the ESR, and would be compatible with the operation capabilities of the EIC complex. 
 
\subsection{The electron linac}\label{sec:electron}

The electron beams of the EIC are generated in an existing polarized source that has demonstrated 90~\% polarized
electron bunches with up to 10~\nC\ charge~\cite{spin-litvinenko-conventional}. The source can be operated with a frequency of 28~\Hz\ generating 1~\nC\ electron bunches.  A Wien filter will rotate the electron spin into the vertical direction, and this polarization is maintained during the entire acceleration process until injection into the ESR.

In the proposed plasma-based injector, longitudinally polarized electrons are accelerated to 150~\MeV\ before being injected into the plasma accelerator. These electrons are extracted from the polarized source and injected into a 500~\MHz\ bunching section followed by a conventional s-band linac, with three accelerating structures of 3~\m\ length spaced by a 1~\m\ drift to allow focusing and correction elements as well as utilities to be placed. 
The RF power is generated in three s-band klystrons and the RF-pulse structure is generated in commercial solid-state modulators. The structures generate an accelerating gradient of 16.6~\MVm.

This pre-injector will provide 1~\nC\ bunches with a frequency of 20~\Hz. The energy spread will be 1\% and the transverse emittance will be at most 10~\um.  The beam is then injected  into the plasma accelerator, as discussed in Section~\ref{sec:plasma}.

\section{Plasma Sections}\label{sec:plasma}
We now turn to the components that are based on plasma.  We begin with a description of the electron bunch acceleration in plasma.

\subsection{Proton-driven Plasma Wakefields}
\label{sec:PlasmaAccStudies}
\subsubsection{Self-modulation stage}

As discussed in Section~\ref{sec:AWAKE}, the long proton bunch delivered from the proton accelerator in Section~\ref{sec:proton} must be conditioned before it can drive wakefields suitable for the acceleration of a witness bunch of electrons.  The plasma electrons react to the space charge of the proton bunch as it passes, resulting in a plasma wave.  In the linear regime, where the amplitude of the wave is small compared to the cold wavebreaking limit $E_c=mc\omega_p/e$, the plasma response can be determined by the Green's function. Here, $\omega_p=\sqrt{ne^2/\varepsilon_0m}$ is the plasma frequency, with $n$ the number density of the unperturbed plasma, $-e$ and $m$ the electron charge and mass, $\varepsilon_0$ the vacuum permittivity and $c$ the vacuum speed of light.  In the co-moving frame, $\tau=t-z/c$, the on-axis accelerating field $E_z$ is then given by:
\begin{gather}
  E_z(\tau)=-\frac{k_p^2}{2\pi\varepsilon_0} \int^\tau_{-\infty} I_\textrm{eff}(\tau^\prime) \cos{\left(\omega_p\left(\tau-\tau^\prime\right)\right)} \D\tau^\prime
\end{gather}
where $k_p=\omega_p/c$ is the inverse plasma skin depth, and $I_\mathrm{eff}$ the effective current, a measure of the wakefield amplitude excited by each slice of the beam~\cite{pwfa-katsouleas-beamloading,pwfa-keinigs-dynamics}:
\begin{gather}
  I_\mathrm{eff}(\tau)=v_b\int_0^\infty r\rho_b(\tau,r)\operatorname{K}_0(k_p r)\D r,
\end{gather}
with $v_b\sim c$ the driver velocity, $\rho_b$ the driver charge density, and $\operatorname{K}_0$ the modified Bessel function of the second kind.  For the large-amplitude wakefields considered in this work, simulations are required to model the nonlinear response of the plasma and the evolution of the proton driver.  However, linear theory remains a useful tool to understand the underlying processes.  Simulations including ion motion were carried out using the particle-in-cell code LCODE~\cite{pic-lotov-lcode,pic-lcode-manual}.

\begin{figure}
    \centering
    \includegraphics[width=1\linewidth]{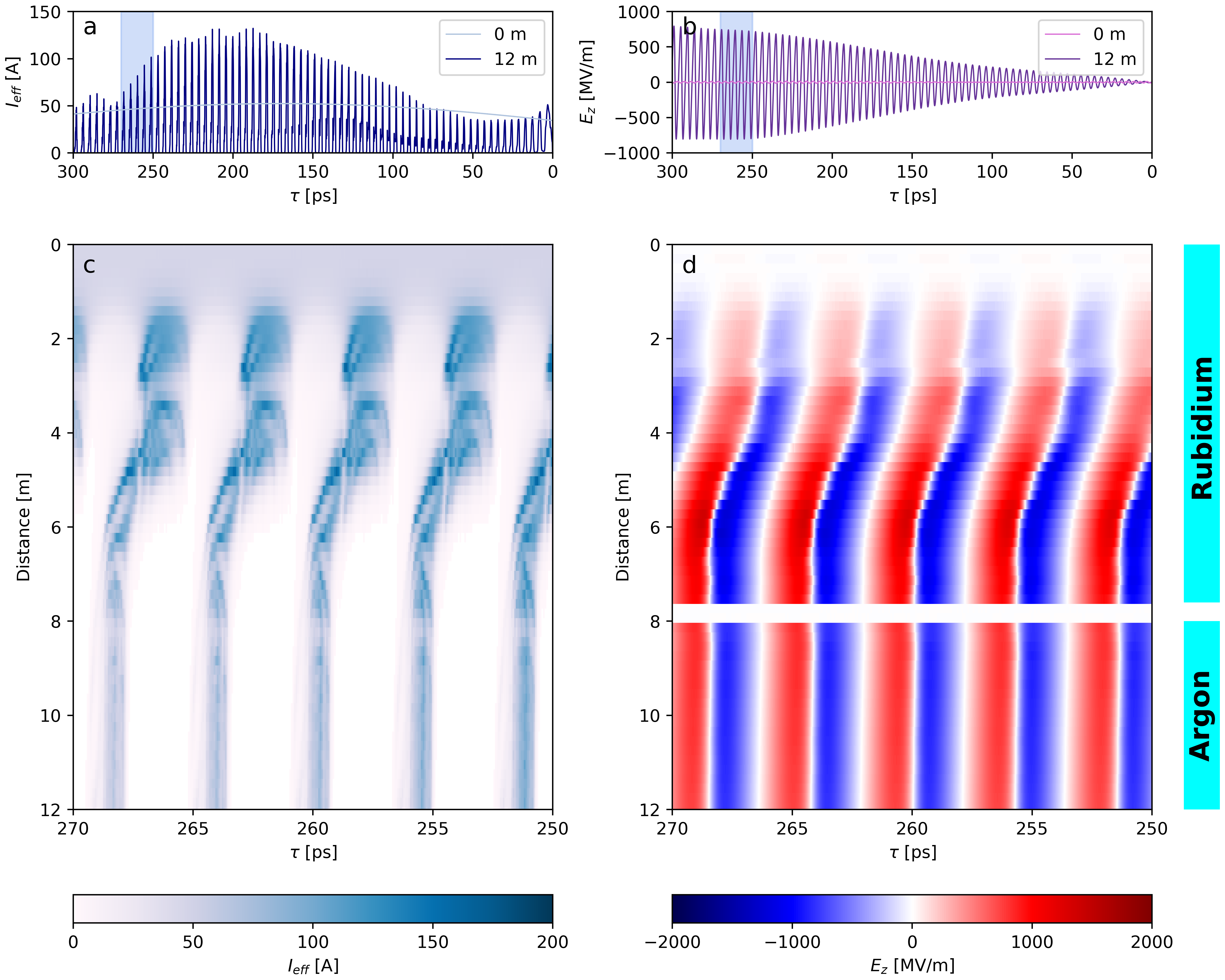}
    \caption{a) The effective current of the proton bunch and b) the associated longitudinal wakefield at the beginning of the first plasma and after 12~\m\ of propagation.  c) and d) The evolution of these quantities over the 12~\unit{\metre}, consisting of the first plasma source and beginning of the second plasma source, showing self-modulation followed by phase stabilization.  For clarity, c and d show a zoomed-in region of $\tau$, corresponding to the shaded areas in a and b.}   
    \label{fig:plasma_modulator}
\end{figure}

The proton beam is focused to a waist size of 200~\um\ at the entrance of the first plasma source consisting of rubidium vapour with a number density of \num{7e14}~\ccm.  The ionizing laser co-propagates with the proton beam, $1 \sigma_z$ (170~\ps) ahead of the bunch centroid. The portion of the beam propagating ahead of RIF is not subject to plasma wakefields, while the portion behind the RIF is periodically focused and defocused, leading to self-modulation.  The RIF provides a fixed point in the co-moving frame from which self-modulation grows, providing a reproducible wakefield phase~\cite{pwfa-batsch-seeded}.

Figure~\ref{fig:plasma_modulator}a shows snapshots of the proton driver as it first enters the plasma and after 12~\m\ of plasma propagation, at which point it is fully modulated.  The wakefield excited by different longitudinal slices of the unmodulated bunch interfere destructively, resulting in low wakefield amplitudes, as shown in Fig.~\ref{fig:plasma_modulator}b.  Although these initial fields are too small to be suitable for acceleration, they are sufficient to periodically focus and defocus the proton bunch over a propagation distance of $\sim2$~\unit{\metre}.  The focused regions of the beam form microbunches, which have a much larger effective current than the defocused regions.  The microbunches are periodic on the plasma frequency, and so their contributions sum coherently, resulting in wakefields of 800~\MVm\ after modulation.

As the wakefields grow, nonlinearities in the plasma response lead to a shift in the resonant plasma frequency.  This results in a phase shift of the plasma wakefields which, if not corrected, leads to the microbunch train being defocused and a decrease in the wakefield amplitude.  A plasma density step early in the self-modulation process allows the wakefield phase to be corrected~\cite{pwfa-lotov-step}.  The result is a train of microbunches which survive for long propagation distances~\cite{pwfa-lotov-beyonddephasing,pwfa-jaworksa-AWAKE4EIC}, allowing the acceleration of a witness bunch to high energy.  For these parameters, a 4\% density step after 0.5~\m\ of plasma was chosen to stabilize the wakefields.

The phase of the microbunch train and the wakefields it excites varies rapidly during self-modulation, as can be seen in Figs.~\ref{fig:plasma_modulator}c,~d, which show the evolution of a section of the proton driver and the associated wakefields.  A 30~\cm\ gap is therefore introduced after 7.7~\m\ of plasma, allowing a witness bunch of electrons to be injected into the phase-stable wakefields for acceleration.

\subsubsection{Acceleration stage}
\begin{figure}
    \centering
    \includegraphics[width=\linewidth]{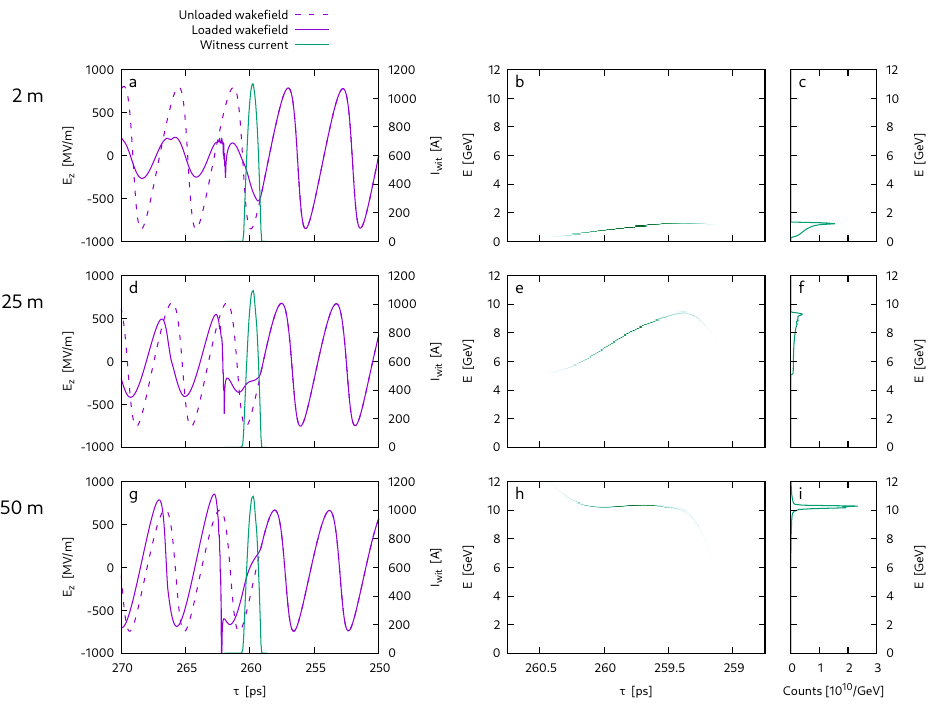}
    \caption{a,~d,~g) the longitudinal wakefield acting on the electron bunch after 2, 25, and 50~\unit{\metre} acceleration, respectively.  The wakefields both in the absence (``unloaded'') and presence (``loaded'') of the 1~\nC\ electron bunch are shown.  b,~e,~h) the corresponding longitudinal phase space and c,~f,~i) energy spectrum.}
    \label{fig:plasma_load}
\end{figure}

Acceleration of the electron bunch takes place in a second plasma with a length of 50~\m.  An argon discharge was chosen for this source, as discussed in Section~\ref{sec:PlasmaSource}.  Figure~\ref{fig:plasma_load}a shows the electron bunch and the longitudinal wakefields acting upon it after 2~\m\ of plasma propagation. Although the modulated proton bunch excites wakefield amplitudes of $\sim800$~\MVm\ (the ``unloaded'' wake), the electron bunch also drives wakefields.  In the accelerating phase, the wakes driven by the proton and electron bunches destructively interfere, with the electron bunch accelerated in the process.

The 1~\nC\ electron bunch charge, necessary for the EIC injector, is much larger than that planned in AWAKE~\cite{pwfa-AWAKE-symmetry}, and so the impact of beamloading is more significant. Beamloading can be seen in Fig.~\ref{fig:plasma_load}a, where the wakefields in the presence of the electron bunch (the ``loaded'' wake) are significantly lower than the unloaded wake. The wakefields become strongly nonlinear after the passage of the electron bunch, consistent with the formation of a blowout~\cite{pwfa-rosenzweig-quasinonlinear}.

The disparate masses of the protons and electrons results in bunches which propagate at different velocities.  The electron bunch is essentially stationary in the light frame $\tau$, while the proton bunch and the wakefields it excites gradually fall backwards.  Injection into the second plasma cell avoids the rapid phase variation which occurs during the formation of the microbunch train, and the use of a density step in the first plasma results in wakefield amplitudes which remain essentially constant over the full acceleration length.  However, the slow variation in the phase of the proton-driven wakefields due to the different bunch velocities remains, as seen by the varying phase of the unloaded wakefields in Figs.~\ref{fig:plasma_load}a,~d,~g, corresponding to 2, 25 and 50~\m\ of plasma propagation.  This dephasing is the main limitation on the acceleration length~\cite{pwfa-jaworksa-AWAKE4EIC}.

Figures~\ref{fig:plasma_load}b,~e,~h show the evolution of the longitudinal phase space.  The combined effect of dephasing and beamloading is that the loaded accelerating field sweeps through the electron bunch, with the head accelerated first, and then the tail. The longitudinal phase space of the electron bunch develops a characteristic `s'-shaped profile during the acceleration, with the phase space rotating over the acceleration length. The  injection position and the electron bunch duration were chosen such that the energy spread, shown in Figs.~\ref{fig:plasma_load}c,~f,~i, is minimized after acceleration, with the head and tail reaching the same energy.

It can be seen from Fig.~\ref{fig:plasma_load}h that at end of the acceleration length, the loaded wakefield acting on the electron bunch head is decelerating, while that acting on the tail remains accelerating.  The net energy transfer from the proton-driven wake to the electron bunch reaches zero, with the amplitude of the plasma wave effectively unchanged by the electron bunch.  However, these final stages of propagation in plasma act to de-chirp the electron bunch, resulting in a narrowly peaked final spectrum shown in Fig.~\ref{fig:plasma_load}i.

The 1~\unit{\nano\coulomb} electron bunch is accelerated to 10~\GeV\ in 50~\m\ of plasma.  The final bunch has a quasimonoenergetic energy spread typical of wakefield accelerators, with 63\% of the bunch charge within $\pm1$\% of the peak after 50~\m\ acceleration.  The tails of the energy spectrum (the ends of the `s') arise due to the longitudinal profile of the electron bunch, and can be removed through proper shaping of the current~\cite{pwfa-meer-beamloading,Lindstrom:2021tkb}.  Although beyond the scope of this work, shaping the electron bunch should also lead to an increase in efficiency, allowing the electron bunch charge to be increased.

The normalised emittance of the electron bunch increases from 10 to 14~\um\ over the full acceleration length, due to the finite time for the electron-driven blowout to form~\cite{pwfa-olsen-inject}.  Further increases in emittance may arise due to misalignment between the proton driver and electron beam~\cite{pwfa_wilson_align}.  However, misalignment will typically be larger in the horizontal plane due to jitter on the proton beam kicker, resulting in an emittance growth predominantly in the horizontal plane.  The emittance of the plasma-accelerated beam is therefore expected to remain below the nominal values of e.g.\ 391/26~\um\ horizontal/vertical normalised emittance for 10~\GeV\ beams given in Table~\ref{tab:ESR-params}~\cite{accel-willeke-eic_cdr}.  After injection into the ESR, the beam emittance is expected to grow until it reaches the nominal values.

\begin{figure}
    \centering
    \includegraphics[width=\linewidth]{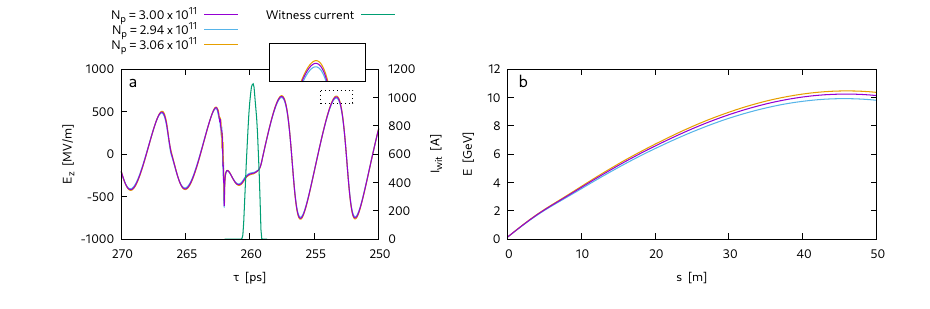}
    \caption{a) the loaded longitudinal wakefield after 25~\unit{\metre} of acceleration for varying drive bunch population.  The nominal case of \num{3e11} protons corresponds to Fig.~\ref{fig:plasma_load}d, above.  b) small fluctuations in the proton beam charge lead to variations of the same order in the mean energy after acceleration.}
    \label{fig:plasma_pop}
\end{figure}

Variations in the proton beam parameters will also impact on the wakefields and the acceleration of the electron bunch.  Changes to the beam population, length and timing will all manifest in a similar way, modifying the initial current profile between the RIF and the electron bunch.  Figure~\ref{fig:plasma_pop}a shows the loaded wakefields after 25~\m\ of acceleration, for the nominal drive beam (equivalent to that shown in Fig.~\ref{fig:plasma_load}d), and for $\pm2$~\% proton beam population.  As can be seen, these variations lead to small changes in the wakefield amplitude, which in turn result in variations in the energy gain, shown in Fig.~\ref{fig:plasma_pop}b.  The change in energy gain is on the same order as the variation in population, with the peak energy varying by $+3.2$ \% and $-2.2$ \%, respectively.

Shot-to-shot variations in electron bunch energy may also arise due to timing jitter between the RIF and the electron beam, or variations in the plasma density.  Corrections are therefore needed to address both the single-shot energy spread and the variation in energy between shots. A method to correct the electron bunch energy after the plasma acceleration stage is discussed in Section~\ref{sec:correct}.

The development of deleterious instabilities, most notably hosing, must also be suppressed over the acceleration length.  Simulation studies show that seeding self-modulation suppresses the growth of hosing~\cite{pwfa-vieira-hosingsuppression}, which is consistent with experiment~\cite{pwfa-clairembaud-reproducibility}.  The proposed scheme also makes use of RIF-seeding~\cite{pwfa-batsch-seeded}, which is inherently symmetric about the bunch, avoiding possible contributions from misalignment~\cite{pwfa_nechaeva_hosing}. 
\subsection{Plasma Acceleration of Polarized Electrons}

\label{sec:PlasmaPolarized}
A key requirement for the EIC electron injector is the preservation of the high degree of polarization provided by the conventional source during the subsequent plasma-based acceleration stage. For externally injected electron beams, this question can be addressed within the framework of spin dynamics in strong electromagnetic fields, governed by the Thomas–Bargmann–Michel–Telegdi (T-BMT) equation:
\begin{align}
 \frac{\D \ssb}{\D t} = - \Omegab \times \ssb \; .
\end{align}
Here, $\ssb$ is the polarization vector and $\Omegab$ the precession frequency:
    \begin{align}
        \Omegab = \frac{e}{mc} \left[ \Omega_B \Bb - \Omega_V \left( \frac{\vb}{c} \cdot \Bb \right) \frac{\vb}{c} - \Omega_E \frac{\vb}{c} \times \Eb \right] \; , \label{eq:prec}
    \end{align}
with $\vb$ the particle velocity and $\Eb, \Bb$ the electric and magnetic field vectors. 
The prefactors in the above equation are given by:
    \begin{align}
        \Omega_B = a + \frac{1}{\gamma} \; , && \Omega_V = \frac{a \gamma}{\gamma + 1} \; , && \Omega_E = a + \frac{1}{\gamma + 1} \; ,
    \end{align}
and depend on the Lorentz factor $\gamma$ and the electron anomalous magnetic moment $a = (g-2)/2 \approx 10^{-3}$, with $g$ the dimensionless magnetic moment of the electron.

In a plasma wakefield accelerator, electrons experience strong transverse focusing fields and longitudinal accelerating fields, which can induce spin precession and lead to depolarization. An overview of the general state-of-the-art is given by Reichwein \textit{et al.} in \cite{spin-reichwein-review}.  Particle-in-cell simulations that include spin tracking  \cite{spin-wu-lwfa, spin-wu-pwfa} show that depolarization effects are dominated by the injection phase, where electrons are still moderately relativistic, $\gamma \ll 1/a$, and may sample rapidly varying electromagnetic fields. In contrast, during the main acceleration stage, where electrons are accelerated from hundreds of MeV to multi-GeV energies and have $\gamma \ge 1/a$, the spin drift is greatly reduced.  For energies approaching the \unit{\tera\electronvolt}-scale, well beyond those considered in this work, radiative depolarization may become important~\cite{pwfa-mathiak-radiative_depolarization}.

The main contribution to spin precession for the acceleration scheme considered here is therefore due to the transverse wakefields acting on the bunch.  However, as the highly relativistic electrons execute betatron oscillations about the beam axis, the direction of the focusing field changes sign, and so the direction of spin precession also oscillates~\cite{spin-vieira}. Since electrons populate all possible betatron phases, this causes a spread in the bunch polarization. For a bunch of radius $\sigma_r$ the polarization spread $\Delta \ssb$ for longitudinally polarized electrons
calculated in \cite{spin-vieira} is:
\begin{equation}
    \frac{|\Delta \ssb|} {s_0} = \frac{1}{8\sqrt{2}} \left( 1 + s_0^2 \right) k_p^2 \sigma_r^2 \gamma a^2 \label{depolarSigma}
\end{equation}
where $s_0=\textbf{e}_z\cdot\ssb_0$ is the initial longitudinal polarization.

A matched electron bunch with normalized emittance $\epsilon_n$ has an equilibrium radius
 \begin{equation}
        \sigma_r =  \left( \frac{2}{\gamma} \right)^{1/4} \left( k_p^{-1} \epsilon_n  \right)^{1/2}. \label{eq:matched}
\end{equation}
Substituting (\ref{eq:matched}) into (\ref{depolarSigma}), we obtain the general expression for the polarization spread of an electron bunch caused by plasma wakefields,
\begin{equation}\label{eq:spin-matched}
    \frac{|\Delta \ssb|}{s_0} = \frac{1}{8} \left( 1 + s_0^2 \right) k_p \epsilon_n \gamma^{1/2} a^2.
\end{equation}

This result has important implications for the present scheme. In the proposed EIC injector, electrons are externally injected from a conventional polarized source with an energy of 150 MeV and are rapidly accelerated to multi-GeV energies within the plasma stage. Since the injection process can be carefully controlled (e.g.\ via phase-stable injection into the wake), the main contribution to depolarization can be minimized. Once the electrons reach relativistic energies, subsequent acceleration to the final energy of 10–18~GeV occurs in a regime where spin precession effects are strongly suppressed.
Taking the proposed EIC injector electron bunch parameters of $\gamma= \num{2e4}$, $\epsilon_n=14~\um$, and $s_0=0.85$, and the plasma skin depth $k_p^{-1} = 200~\um$, we obtain the expected polarization spread caused by the plasma wakefield to be 
${|\Delta {\bf s|}} / {s_0} \approx \num{2.1e-6}$.

As discussed in Section~\ref{sec:PlasmaAccStudies}, if the electron bunch injected into the plasma is mismatched or misaligned, the emittance will rapidly grow until a new equilibrium is achieved.  However, since this emittance growth is limited to the initial stages of acceleration, when the electron energy is low, the corresponding betatron period is short and the integrated spin precession over a half oscillation remains small.  After a new matching equilibrium is reached, the polarisation spread scales linearly with the new emittance, as in Eq.~(\ref{eq:spin-matched}).  Even a large increase in the initial emittance will therefore not lead to significant depolarization.  The beamloading observed in simulation, which reduces $E_z$, and the longitudinal phase-space rotation which occurs during acceleration, do not influence the spin precession of the longitudinally polarized beam as the contribution vanishes for $v \parallel E$.

The plasma wakefield accelerator therefore avoids the potential for depolarization which may occur in the booster ring in the baseline design. It should be noted that conservation of polarization for an externally injected beam in plasma has yet to be experimentally demonstrated, although the underlying physics dictates that it will be.  Preliminary experiments have been carried out for laser-accelerated ion beams~\cite{Zheng_polarized}.

Overall, both analytical considerations and numerical studies indicate that polarization preservation in plasma wakefield acceleration is robust for externally injected, highly relativistic electron beams. Therefore, the high initial polarization ($\sim 85\%$) provided by the source can be maintained throughout the plasma acceleration stage.  The resulting average polarization which can be achieved in the ESR is discussed in Section~\ref{sec:topup}.
 
\subsection{Plasma cell concepts}
\label{sec:PlasmaSource}

The proposed injector 
closely follows the baseline design for AWAKE Run~2d~\cite{pwfa-AWAKE-symmetry}
and employs two distinct plasma sources. 
As discussed in Section~\ref{sec:PlasmaAccStudies}, the first plasma induces self-modulation of the proton bunch into a train of microbunches that resonantly drive a wakefield in the second plasma section, which then accelerates a witness electron bunch.  
A small gap between the two plasmas enables on-axis injection of the electron bunch into the wakefield using a magnetic dipole, which is essential for preserving beam quality.

The phase of the accelerating wakefields excited by the proton bunch is achieved using a RIF synchronized with the electron bunch, which constrains the first plasma to a laser-ionized alkali source. 
Uniform-density plasmas with lengths of $\sim \SI{10}{\meter}$ can be produced via field ionization 
using a short-pulse, high-intensity laser with a long focal length, with the vapour density precisely controlled through the temperature of the source vessel.

The alkali vapour plasma source developed for AWAKE~\cite{Plyushchev:2015kta} uses rubidium vapour. The choice of rubidium is motivated by its low first ionization potential, the sufficiently large gap to the second ionization potential, its relatively high ion mass, and the comparatively low temperature required for the vapour to reach the target densities (\SIrange{150}{250}{\celsius}). 
Although the repetition rate considered here (\SI{20}{\hertz}) is significantly higher than that of AWAKE (\SI{0.05}{\hertz}), it remains compatible with the technological solution developed for this scheme.

The proposed RIF scheme directly impacts the beamline layout, as long-focal-length optics ($\sim \SI{50}{\meter}$) are required to produce plasma lengths of up to \SI{10}{\meter} using a short-pulse ($\sim \SI{50}{\femto\second}$) CPA laser with an energy of $\sim \SI{200}{\milli\joule}$. This constrains the last mirror in the laser beamline to be placed at $\sim \SI{20}{\meter}$ upstream of the first (rubidium vapour) plasma section, while leaving sufficient space to bend the proton beam trajectory towards the laser propagation axis using a dipole magnet.

The controllable plasma density step, adjustable in amplitude and axial position and required to maintain the accelerating gradient, can be implemented in the rubidium vapour source by imposing a corresponding temperature step using independently controlled heating sections along the plasma cell.

\begin{figure}
    \centering
    \includegraphics[width=1.0\linewidth]{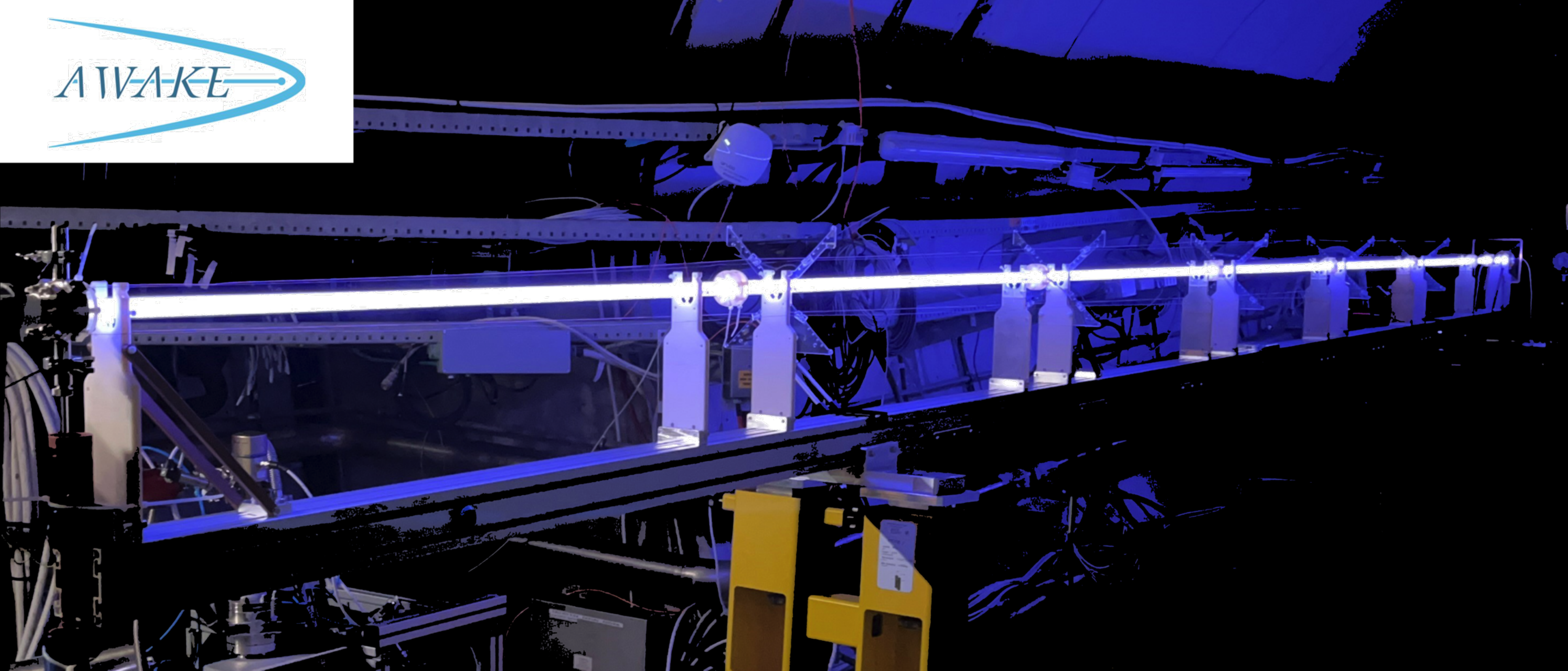}
    \caption{Discharge plasma source (DPS) operating in the AWAKE experiment. The image shows the DPS configured as a single plasma column of \SI{10}{\meter} length, with plasma generated in argon at \SI{8}{\pascal} using a \SI{350}{\ampere} peak-current heating pulse (courtesy of A. Sublet and the AWAKE team).}
    \label{fig:DPS_tunnel_pic}
\end{figure}

The second plasma, used for acceleration, 
has an electron density matching that of the preceding modulator plasma and extending over the required acceleration length (\SI{50}{\meter} in this case). 
Generating such a long plasma using laser field ionization of an alkali vapour would require a very high-power laser and impractically long focal-length optics. Therefore, closely following the AWAKE Run 2d baseline design, we propose to use a length-scalable plasma source capable of producing a long preformed plasma column.

Two approaches are currently under development: helicon discharges
\cite{Buttenschon:2018_HPS1,Granetzny:2025_HPS2} 
and direct current discharges (DPS) 
\cite{Lopes:2026_PPCF_meterScale,plasma-torrado-discharge,torrado:2025_JPhys_doPulse_Scala,plasma-amoedo-time_resolved_dps}.
We propose to use a DPS for the second plasma since it has already been demonstrated in AWAKE, generating plasma columns with lengths up to \SI{10}{\meter} 
\cite{torrado:2025_JPhys_doPulse_Scala} 
and sufficient uniformity to self-modulate SPS \SI{400}{\giga\electronvolt} proton bunches, with results similar to the standard RIF method 
\cite{pwfa-amoedo-dischargedemonstration,turner:2025_PRL_IonMass}
(excluding phase-locking, since this is a preformed plasma and there is no SMI seeding). A photograph of the DPS generating a \SI{10}{\meter} long plasma in argon is shown in Fig. \ref{fig:DPS_tunnel_pic}.

DPS plasmas reach the target electron density using a short-pulse (a few microseconds), high-current density (\(\approx\)\SI{100}{\kilo\ampere\per\square\centi\meter}) discharge between electrodes located at the extremities of a long dielectric tube filled with  low-pressure gas (\SIrange{2}{50}{\pascal}). A low-jitter ignition of these short-pulse arc discharges is crucial for synchronizing the plasma discharge with the particle beams. We adopt a double-pulse ignition-heating approach~\cite{plasma-torrado-discharge} 
in which both pulses are direct-current arc discharges between a cold cathode and an anode immersed in the gas.

The initial discharge (ignition) is driven by a fast-rise-time high-voltage pulse (up to $\approx$ \SI{50}{\kilo\volt}) generated by a flyback transformer that reliably initiates the plasma via cold electron emission at the cathode enhanced by the intensified electric field between the cathode and the current-return tube shield. 
This produces a diffuse and uniform plasma with a low ionization fraction (currents in the range \SIrange{10}{50}{\ampere}) along the tube.
Higher ionization fractions ($\sim$ \SIrange{10}{60}{\percent}, depending on the gas and pressure) are subsequently achieved using a lower-voltage (\SIrange{4}{10}{\kilo\volt}) efficient capacitive discharge (heating pulse).

The choice of gas is mainly constrained by the ion mass required for negligible ion motion that would otherwise reduce the wakefield amplitude within the proton bunch duration. 
This restricts the usable gases to those with ion masses greater than or equal to that of argon 
\cite{vieira:2012_PRL_IonMass,turner:2025_PRL_IonMass,pwfa-walter-ionmotion}.

\begin{figure}
    \centering
    \includegraphics[width=1.0\linewidth]{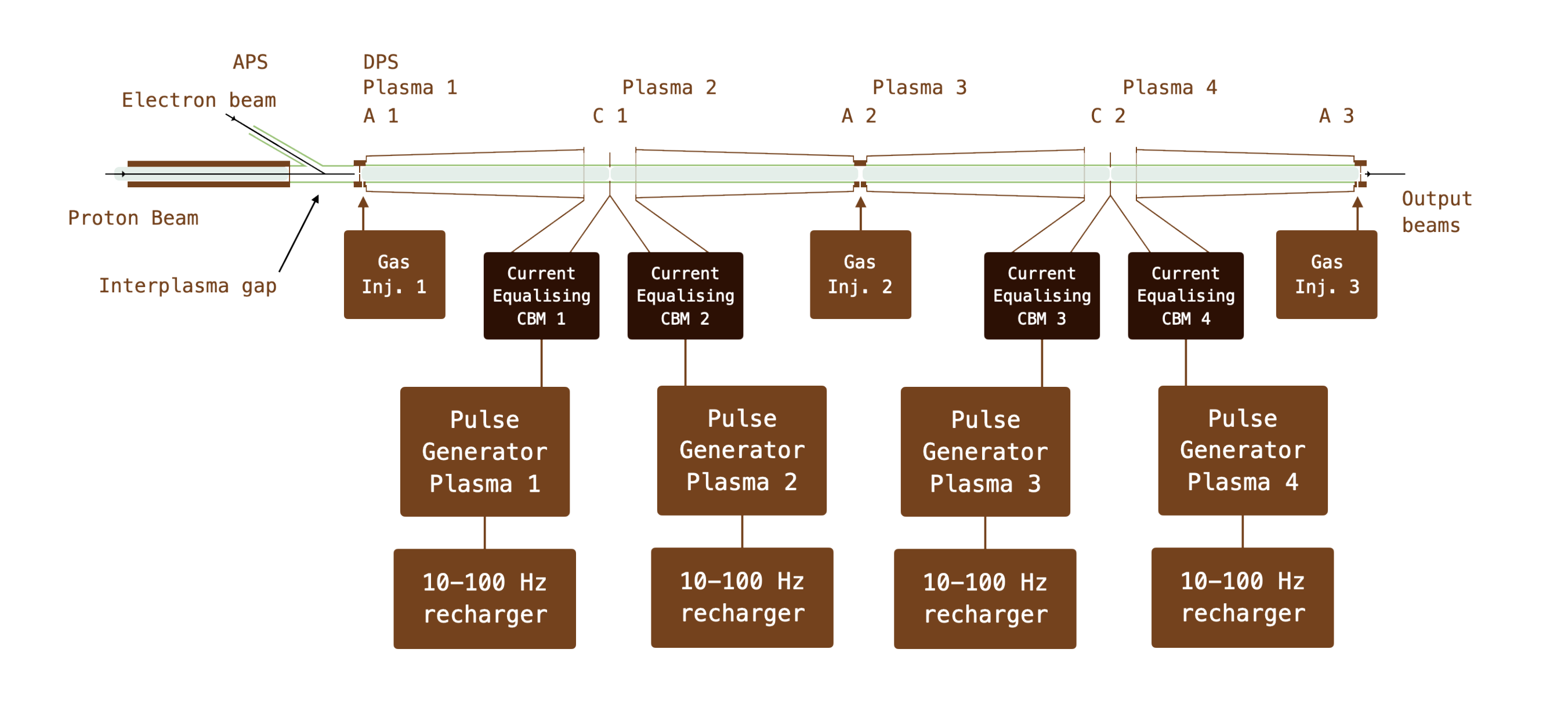}
    \caption{Simplified scheme of the DPS consisting of four equal plasmas with a total length of \SI{50}{\meter} (not in scale). The scheme also shows the end of the laser-ionized alkali plasma source and the inter-plasma gap - used for electron injection into the wakefield generated by the self-modulated proton bunch in the DPS plasma. This scheme can be used to produce arbitrarily long plasmas by joining additional double plasma sections. Each pulse generator produces the consecutive pulses to generate each plasma, with current-equalizing magnetic chokes used to force equal currents in each plasma. The entire plasma source is required to be electrically isolated from adjacent metal parts with the grounding of each module to be made inside the respective pulse generator.}
    \label{fig:DPS_scheme}
\end{figure}

The DPS has been tested for plasma lengths of up to \SI{10}{\meter} and is expected to perform well for longer lengths. However, since the heating pulse voltage needs to be proportional to the plasma tube length, significantly longer plasmas will require the combination of multiple discharges. The first step in length-scaling consists of producing two plasmas in a tube with two oppositely directed discharges originating from a common cathode, located at the tube midpoint and propagating towards anodes at the tube extremities.  The currents in both plasmas are forced to be equal by a set of magnetic chokes that suppress current imbalance. Arbitrarily long plasma lengths can be achieved by combining these double plasma sections as schematically shown in  Fig. \ref{fig:DPS_scheme} for a combination of four plasma sections.

A demonstration of a \SI{12}{\meter} long plasma consisting of two double plasma sections (four plasma discharges in total, in a setup similar to the one shown in  Fig. \ref{fig:DPS_scheme}) has recently been performed. Preliminary results indicate that this method generated equal currents in the four plasma sections (to within \SI{1}{\percent}), essential for uniform length-scalable plasmas. If required, the fine tuning of the currents can be achieved by modifying the charging voltages in each pulse generator.

When needed, ionization fractions close to 100~\% can be produced by combining a higher heating discharge current with a solenoid magnetic field ($\sim$ \SIrange{0.2}{1.0}{\tesla} depending on gas and plasma tube diameter). 
These magnetic fields are not expected to affect the wakefield and the acceleration process but instead mildly confine the plasma, reducing the energy transport to the tube walls. 
This results in an increase in plasma temperature and, consequently, the ionization fraction.
For example, argon plasmas ionized above \SI{90}{\percent} can be obtained with a current density of \SI{80}{\ampere\per\square\centi\meter} and a solenoid magnetic field of \SI{0.7}{\tesla}. 
Using a high ionization fraction  plasma can lead to increased plasma uniformity if a plateau in ionization fraction versus density current is reached, since this reduces the dependence of the plasma density on the inner tube diameter tolerance. In this application, since the repetition rate allows for plasma full recombination~\cite{pwfa-gessner-recovery}, we aim for low current densities and ionization fractions ($\sim \SI{20}{\percent}$ in argon). This avoids the use of the solenoid magnetic field but imposes a strict tolerance of the dielectric tube diameter to ensure the required current density uniformity.

Although the proposed plasma sources are now relatively well tested, the detailed implementation of the two plasma sections scheme is still in the design phase within the AWAKE experiment. One of the main design priorities is the reduction of the interplasma gap length, which is required to enable on-axis injection of the witness electron beam into the wakefield of the second plasma. Longer free-propagation distances of the self-modulated proton bunch lead to a reduction in wakefield acceleration efficiency.

This interplasma region is densely populated with critical components, including the electron injection dipole, the downstream end of the alkali plasma source (comprising a laser beam dump and diagnostics of proton bunch self-modulation), and the upstream end of the second plasma (including a discharge anode that must be electrically isolated from surrounding components), as well as beam diagnostics for quality assessment and for alignment and synchronization of the proton and electron beams.

Precise temperature control is required in both plasma cells to ensure density uniformity and to generate the controlled density step in the first plasma cell.
The proposed EIC injector is expected to operate in burst mode, with significant energy intermittently transferred from the proton beam to the plasma and surrounding structures. This may induce plasma density fluctuations due to non-uniform heating, cooling transients, and mechanical perturbations of the discharge tube. Therefore, a detailed thermal design is required for each implementation, taking into account repetition rate, duty cycle, beam energy, and energy transport to the walls. 
The thermal stability of the vapour or gas, which translates into plasma density uniformity, is especially important when using long proton bunches that self modulate into a large number $N_{mb}$ of microbunches, with $N_{mb} \approx k_p L_b / 2\pi$, where $L_b$ is the bunch length.
In order to maintain the quality of the accelerated electron bunch, the variations of the wakefield phase in the tail of the bunch must be kept below a quarter wakefield wavelength, 
corresponding to
$\smash{\Delta k_p / k_p \leq 1/(4N_{mb})}$, 
which translates to $\smash{\Delta n_e / n_e \leq 1/(2N_{mb}) \approx \num{1.6e4} / (L_b n_e^{1/2})}$. 
For $L_b = \SI{0.05}{\meter}$ and $n_e = \SI{7e14}{\per\cubic\centi\meter}$ we obtain $\Delta n_e / n_e \leq \SI{1.2}{\percent}$.

The higher repetition rate, compared with AWAKE, imposes two additional changes on the DPS modules. 
The plasma ignition pulse, which in AWAKE is a sharp rise-time pulse with a long decay (triangular waveform), 
must now have short rise and fall times, since long pulses lead to longitudinal density modulations that may affect the plasma uniformity at \SI{20}{\hertz}. 
This problem can be avoided by changing the ignition circuit from a flyback transformer to a Marx bank topology. 
The second change is related to plasma ion drift. The plasma ions are accelerated along the tube from anode to cathode during the discharge. Keeping the heating pulses short (few microseconds) reduces the drift to $\sim \SI{1}{\centi\meter}$. Although this drift is negligible for a single pulse, at \SI{20}{\hertz} it may result in gas accumulation/rarefaction at cathode/anode regions. This problem can be circumvented, as in argon ion lasers, by a return tube that maintains the pressure equilibrium between cathodes and anodes. Alternatively, the ion drift can be compensated by using a second high-current pulse, with opposite polarity to the heating pulse, launched immediately after the end of the heating pulse and the beam propagation. This ion drift mitigation pulse in practice duplicates the plasma heating circuit while reducing the complexity of the plasma tube geometry by eliminating the gas return tube. 

The upstream interface of the second plasma is critical for the quality of the accelerated beam. A thin metallic foil (used as the first discharge anode) can confine the gas, leading to improved density stability and plasma uniformity. However, propagation of the relatively low-energy witness electron beam through the foil may degrade beam quality. In that case, the foil must include an aperture for beam injection, and the gas injection system must compensate for the resulting gas losses. Differential pumping would then be required to maintain the upstream vacuum.

The baseline parameters of both plasma sections are summarized in Table \ref{tab:plasma-params}.

\begin{table}[hbpt]
\centering
\begin{tabular}{|l|c|c|}

\hline  
Parameter                           & Units                                      & Value       \\
\hline 
First plasma length (RIF)           & \unit{\meter}                              & 7.7         \\
Second plasma length (DPS)          & \unit{\meter}                              & 50          \\
Plasma density                      & \unit{\per\cubic\centi\meter}              & \num{7e14}  \\
Density uniformity/reproducibility  & \unit{\percent}                            & $\leq$ 1.2  \\
DPS tube inner diameter             & \unit{\milli\meter}                        & 25          \\
DPS plasma sections                 & -                                          & 4           \\
DPS plasma section length           & \unit{\meter}                              & 12.5        \\
DPS heating peak current            & \unit{\ampere}                             & 800         \\
\hline 
\end{tabular}
\caption{Design parameters of the two plasma sources.}
\label{tab:plasma-params}
\end{table}

\section{Injection into the Electron Storage Ring}\label{sec:TuneandInject}
\subsection{Energy correction}\label{sec:correct}
Existing results~\cite{pwfa-meer-beamloading,Lindstrom:2021tkb} indicate that proper shaping of the electron witness bunch (triangular with sharp rising edge) can flatten the wakefields in the region of the electron bunch and achieve nearly uniform accelerating fields.  Energy spreads of $< 1$~\% should be possible with this witness bunch shaping.  The energy spread and centroid of the energy distribution will be affected by timing jitter of the injected electron bunch relative to the wakefield, and by variations in the wakefield amplitude.  As discussed, the electron bunch is synchronized with the ionizing laser pulse.  The RIF generated by this laser pulse initiates the modulation process and determines the timing of the generated wakefields.  Based on expectations developed over the course of the AWAKE project, we anticipate timing jitters between the wakefields and electron bunch of $<50$~\fs.  This phase jitter will result in \%-level absolute energy jitter of the electron bunch.  Variation in the proton-bunch parameters may also result in \%-level energy jitter, as discussed in Section \ref{sec:PlasmaAccStudies}.

The ESR beam is stable for up to $1$~\% energy oscillations \cite{Nosochkov:2024stj}.  However, the transverse dynamic aperture of the ring must accommodate the injection oscillations which are associated with beam accumulation.  Since the transverse dynamic aperture for the ESR approaches zero at the energy acceptance level of $\Delta E/E \simeq 1\, \%$, the energy spread and energy jitter of the injected electron bunch must be significantly below this threshold.  Thus, a scheme for correcting the energy spread and offset of the injected bunch is necessary.

This energy correction could be realized as a chicane with an $R_{56}=-0.131$~\m, e.g., four $12$~\Tm\ bending dipoles producing millimetre-scale path length differences over $\sim 40$~\m\ for energy variations on the \% level.  These path length differences would change the phase of the electrons entering an RF accelerating section.  Since the 90~\um\ electron bunch length after plasma acceleration is much shorter than the change in path length introduced by dispersion, the energy-position correlation is dominated by the $R_{56}$ of the chicane, allowing the nonlinear ``s''-shaped phase space to be corrected.  The RF could be implemented as an X-band structure, offering the dual benefits of high gradient and short wavelengths.

More precise values on the required distances, bending fields and RF voltage will require an investigation into the achievable energy spread and jitter for the electron bunches exiting the plasma section.  As discussed in Section~\ref{sec:PlasmaAccStudies}, percent-level variations in the proton bunch charge give rise to energy variations of similar magnitude. The total RF correction required depends on the energy spread averaged over many shots, including all sources of error.

\subsection{Top-up Injection Scheme}\label{sec:topup}
At an energy of $E=\SI{10}{\giga\electronvolt}$, the ESR in high luminosity operation stores $N_b=1160$ bunches with a charge of $Q_b =28$~\nC. 
The electron bunch charge demonstrated in the plasma acceleration simulations in Section~\ref{sec:PlasmaAccStudies} is only $Q_b = 1$~\nC.
Thus the required electron bunch charge needs to be accumulated in the ESR.

For each proton driver cycle, $N_b= 1160$ drive bunches are provided. These bunches are extracted within a time that is short compared to the driver cycle time, with each drive bunch accelerating a 1~\nC\ electron bunch. The finite electron lifetime in the ring means that somewhat more than 28 accumulations are required in order to reach 1160 electron bunches of 28~\nC.  
Accumulations are spaced in time by 11~\mins, and so it takes more than $28\times\, 11\,\mins = 5\, \hrs$ to complete the injection process. After the full bunch charge has been reached, it is maintained by adding an appropriate amount of charge in each refill cycle until the end of the storage period.  The length of the store is limited by the time at which the proton bunches in the HSR are deteriorated to the extent that a new proton fill needs to be injected and accelerated in the HSR. 

The electron bunch charge in the ESR is:
\begin{equation}
Q_b(t+\tau_\mathrm{fill}) 
= Q_b(t) \exp{
\left(
\frac{ -\tau_\mathrm{fill}} {\tau_e} 
\right) }
+\delta Q 
\end{equation}
where $t$ is the time since the most recent injection, $\tau_\mathrm{fill}$ is the time between subsequent accumulation steps, $\tau_e$ is the electron beam lifetime and $\delta Q$ the electron charge added each accumulation step.  The maximum charge supported by the plasma accelerator is injected each accumulation cycle until the full bunch charge has been reached.  After this point, only the charge loss since the last fill need be injected, $\delta Q = Q_b \exp(-t/\tau_e)$.

The polarization $P$ of the electron beam is another important performance parameter. The beam is injected with an initial polarization of $P_0 = 85\,\%$. The polarization of the stored beam without accumulation would be subject to depolarization with a decay constant of $\tau_p$ until the asymptotic value of $P_{\infty}$ is reached:
 \begin{equation}
P(t) = P_{\infty}+(P_0-P_{\infty}) 
\exp{\left(-\frac{t}{\tau_p}\right)}
 \end{equation}
 The asymptotic polarization is the equilibrium of the Sokolov-Ternov polarization due to the average emission of synchrotron radiation photons and the 
 depolarization due to the stochastic nature of the emission of photons and corresponding sudden changes in the orbit of the electrons.
 The polarization lifetime $\tau_p$ depends on the asymptotic polarization of an ideal accelerator ring $P_\mathrm{BKS}$ and the corresponding polarization  time $\tau_\mathrm{BKS}$, $\tau_p = \tau_\mathrm{BKS} P_{\infty}/P_\mathrm{BKS}$. The polarization during the process of charge accumulation and charge maintenance is
 then given by the following
 recursion formula:
 \begin{equation}
P(t+\tau_\mathrm{fill}) =
\frac{
\left( 
P_{\infty} + (P(t)-P_{\infty})  
\exp{\left( -\frac{\tau_\mathrm{fill}}{\tau_p}\right)}
\right)
 Q_b(t) \exp{\left( -\frac{\tau_\mathrm{fill}}{\tau_e}\right) }
+P_0 \delta Q
}
{Q_b(t) \exp{\left(-\frac{\tau_\mathrm{fill}}{\tau_e}\right)}+\delta Q}
\end{equation}
The parameters that govern the charge accumulation and the polarization development are given in Table~\ref{tab:polarization-development}.  The exact polarization lifetime depends on the detailed optics of the machine, including alignment.

\begin{table}
    \centering
    \begin{tabular}{|l|c|c|c|}
\hline
       Parameter  & Units  & \multicolumn{2}{|c|}{Value} \\\hline
    Electron energy          & \GeV &  10 & 18 \\
\hline
    Maximum bunch charge in ring     & \nC & 28  & 10.7  \\
    Number of bunches in ring, $N_b$       &  & $1160$ & $290$ \\
    Maximum plasma accelerator bunch charge, $\delta Q$ & \nC & 1 & 0.64 \\
    Average time between accumulations   & \mins    & 11 & 2.75 \\
    Electron beam lifetime   & \mins & 600 & 40.5 \\

    Initial polarization $P_0$                     & \% & 85    &  85  \\
    BKS polarization, $P_\mathrm{BKS}$                & \%    & 85.91 & 81  \\
    BKS polarization time, $\tau_\mathrm{BKS}$             & \mins & 699 & 33 \\
    Asymptotic polarization in ring, $P_{\infty}$             & \%    & 50  & 50 \\
    Polarization lifetime, $\tau_p$                 & \mins  &  406 & 20.4 \\
    Beam storage time                 & \mins & 600 & 600 \\
\hline
    \end{tabular}
    \caption{The relevant parameters achieved in the current version of our scheme for charge accumulation and polarization development for ESR operation at 10~\GeV\ and 18~\GeV.}
    \label{tab:polarization-development}
\end{table}

 \begin{figure}
    \centering
    \includegraphics[width=0.8\textwidth]{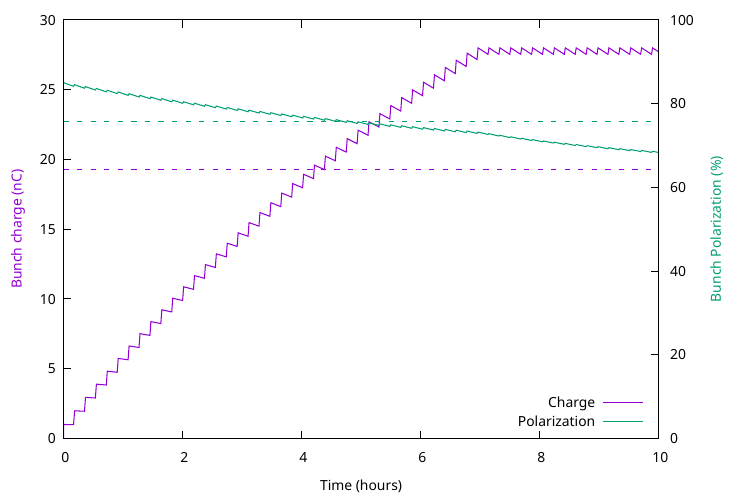}
    \caption{Accumulation of bunch intensity and corresponding polarization development over the store time of $\tau_{store} = 10 \, \hrs$ at 10~\GeV. The average values over the store time are shown by dashed lines, with an average bunch charge of 19.3~\nC, and an average polarization of 75.8~\%.}
    \label{fig:10GeV-acc}
\end{figure}

The corresponding ESR bunch charge accumulation and polarization development for the 10~\GeV\ case are shown in Fig.~\ref{fig:10GeV-acc}.  Note that after the desired electron bunch charge is achieved, the injected charge is reduced so as not to exceed this value. The average bunch charge is 19.3~\nC,  which is 69\% of the EIC baseline design of 28~\nC. The average polarization is 75.8\% which exceeds the EIC performance goal by 8.2~\%.

\begin{figure}
    \centering
    \includegraphics[width=0.8\textwidth]{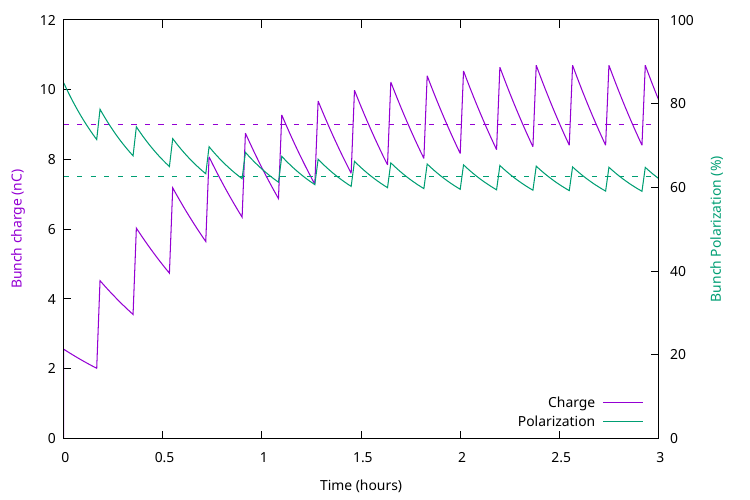}
    \caption{Accumulation of bunch intensity and corresponding polarization development at 18~\GeV. The average values over the full store time $\tau_{store} = 10 \, \hrs$ are shown by dashed lines, with an average bunch charge of 9.0~\nC\ and an average polarization of 62.7~\%.}
    \label{fig:18GeV-acc}
\end{figure}

For 18~\GeV\ operation, only 290 bunches are stored in the ESR, and so each cycle of 1160 drive bunches allows four accumulations per ESR bunch, i.e. four plasma-accelerated bunches injected within the 1-minute injection window every 11 minutes.   The bunch charge in the ESR is also reduced from the value at 10~\GeV.  However, in order to reach 18~\GeV\ in the same length of plasma, the electron bunch charge from the plasma accelerator must also be reduced in order to decrease the beamloading.  We estimate the attainable bunch charge by interpolating between the energy gain for the 10~\GeV\ bunch and the integrated field in the unloaded case~\cite{pwfa-jaworksa-AWAKE4EIC}, and attain a value of 640~\pC. Although not discussed here, operation at 5~\GeV\ would allow more charge to be accelerated.

Unlike operation at 10~\GeV, where the main challenge is achieving the high beam current, the main limitation at 18~\GeV\ is the short polarization lifetime, much shorter than the natural beam lifetime of $\sim600$ minutes.  In this regime, injected bunches would depolarize over timescales much shorter than the rate at which they are replenished with freshly injected polarized electrons.  We therefore artificially reduce the beam lifetime, which could be achieved using, e.g.\ a septum, such that the beam and polarization lifetimes are comparable.  This allows the average polarization to be increased at the expense of the average bunch charge.  We note that the total beam energy removed from the ring is comparable to that using the swap-out method in the EIC baseline.  Keeping in mind that the effective luminosity scales linearly with bunch charge but quadratically in polarization, we choose a beam lifetime of 40.5~\mins.  The evolution of the ESR bunch charge accumulation and polarization development at 18~\GeV\ over the first three hours of a store are shown in Fig.~\ref{fig:18GeV-acc}.  The average bunch charge over the full 10 hours is 9.0~\nC, 84\% of the EIC baseline design, with an average polarization of 62.7~\%, which is 90~\% of the EIC performance goal of 70\%.

The studied injection scheme is close to the EIC performance requirement but these results are achieved only for antiparallel spin w.r.t.\ the guide field.  Both longitudinal spin orientations are necessary at the interaction point in order to suppress systematic errors, achieved in the baseline design through bunch-to-bunch polarization reversal.  However, since the polarization lifetime for spin parallel to the guide field is significantly lower~\cite{EIC_Xu_inject}, in the top-up scheme we instead envisage switching the polarity of the electron spin rotators when beginning a new beam store.  These are used to convert the vertical spin of the electron bunch into longitudinal spin before the IP (and vice versa after collisions), so switching polarity allows both spin orientations at the IP to be achieved (albeit alternating every beam store) while maintaining antiparallel spin in the ring.

While these results fall somewhat short of the EIC performance requirements, it should be pointed out that there is quite some room for further improvement of the parameters of the plasma wakefield acceleration and an increase of the charge of the witness bunch appears to be an achievable goal.  The calculations of average intensity and average polarization have been repeated for an increased charge of the accelerated electron beam.  At 10~\GeV, accelerating 2~\nC\ of charge per drive bunch allows the full charge of 28~\nC\ to be reached after 2.56~\hrs\, with the corresponding average bunch charge increasing to 24.1~\nC, which is 86\% of
the performance goal. This is a significant improvement compared to the case with only 1~\nC\ charge of the witness bunch. The average polarization of 74\% remains above the performance goal.

In the 18~\GeV\ case, doubling the electron charge per drive bunch to 1.28~\nC\ shifts the desired ratio of bunch to polarization lifetimes.  Again optimizing $Q_bP^2$ gives a beam lifetime of 20 minutes, resulting in an average ESR bunch charge of 8.3~\nC, 78\% of the baseline value, while increasing the average polarization to 68 \%.  It would also be possible to dynamically vary the electron beam lifetime, using a long lifetime during accumulation and a reduced lifetime once the full bunch charge has been reached. We therefore expect that further improvements may be possible.

\section{Potential layout}\label{sec:PotentialLayout}

\begin{figure}
    \centering
    \includegraphics[width=1.0\linewidth]{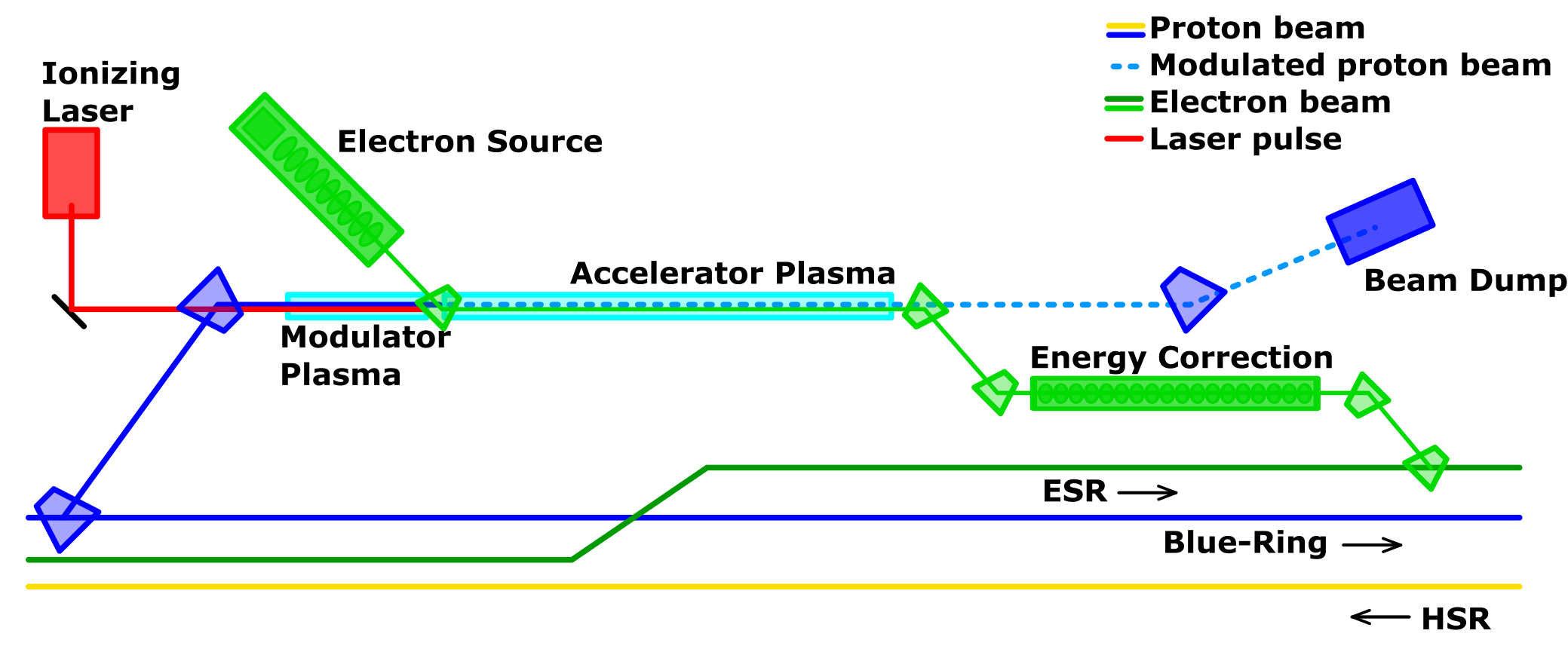}
    \caption{Possible Layout: not to scale. Outline of the necessary beams is shown: proton beam in blue, electron beam in green, laser beam in red. The proton beam self-modulates while passing through the first plasma cell. }
    \label{fig:layout}
\end{figure}

A schematic of the different elements is shown in Fig.~\ref{fig:layout}.  
Proton drive bunches (blue line) are extracted from the second proton ring and transported to the plasma acceleration section. As mentioned in Section \ref{sec:AWAKE}, the bunch has to be modulated, before it can be used for acceleration. The drive-beam modulator consists of a 7.7-metre-long laser-ionized plasma (shown in light blue). A high-power laser pulse (red line) will need to be co-propagated with the incoming proton bunch, to create the plasma and seed the self-modulation process. At the exit of the modulator, a laser dump will be installed to stop the laser pulse. 
After the modulator plasma there is a $30$ cm gap, which provides the necessary space for beam merging, and on-axis electron injection. Electron witness bunches (green line) are generated by a polarized electron source, and pre-accelerated to $150$ MeV using a conventional linac. The beam is then transported to the plasma acceleration section, and injected into the second plasma in the $30$ cm gap. The accelerator cell uses a discharge plasma source, and is expected to be approximately $50$ m long. 

Following acceleration, the proton drive beam is separated from the electron beam, and directed to a proton beam dump. The accelerated electron bunches will be transported through an energy spread correction section, where their energy spread will be reduced to below $1\%$ to match the energy acceptance of the storage ring. The electron beam would then be injected into the ESR. The plasma cells occupy a total of $58$ m, leaving space for electron beam injection, diagnostics, and energy correction. 

The compatibility of such a layout with existing infrastructure will require a detailed investigation.  At present, it appears that an available space of $\approx 200$~m will be sufficient to accommodate the different components.

\section{Discussion and conclusion}\label{sec:conclusion}

In this paper we outline the components and expected performance of an electron injector for the EIC based on proton-driven plasma wakefield acceleration.  
The results of this study indicate that the baseline design parameters of the EIC electron injector are within reach for such a scheme.  We find that acceleration of an electron bunch of $1$~\nC\ (\num{6e9} electrons) to $10$~GeV is possible with a proton bunch of \num{3e11} protons of $275$~GeV.  Reaching higher electron energies would require a reduction in the electron bunch charge, but we note that operation of the electron storage ring at higher energies also requires the beam charge to be reduced due to synchrotron radiation energy losses.  Injecting the accelerated electron bunches into the storage ring at an average rate of $>1$~Hz is feasible when extracting proton drive bunches at tens of \Hz\ and refilling the proton ring within $11$~minutes.  Tens of \Hz\ operation is not expected to be a problem for the plasma sections or for the polarized electron source.  Topping-up the electron bunches in the storage ring at this rate will allow for a high average polarization ($\geq  70$~\% at 10~\GeV) which is critical for the EIC physics program.  The proposed layout indicates that a 200~\m\ straight section would be sufficient for the scheme.  

A number of aspects still need to be verified, such as how best to achieve the $<1$~\% energy spread required for injection into the electron storage ring, including the deviation from the nominal energy, after acceleration.  Achieving this low level of energy spread would likely require the development of a novel electron linac capable of producing tailored longitudinal profiles for injection into the plasma. It would also likely require the development of a final corrector section based on standard RF technology which would be located after the plasma acceleration system to centre the electron bunch energy on the desired energy.  These two RF sections need careful design and detailed studies.  A more detailed study of tolerances is required in order to evaluate the total accelerating voltage needed for producing and shaping the electron bunch before the plasma acceleration and for correcting the energy after the plasma.  We note that the total RF needed is expected to be below that of the 750~\MeV\ linac used in the baseline EIC design for the injection into the booster ring.  As the booster ring for the electron bunches would not be necessary, this scheme could prove substantially less costly, provided a suitable source of high-energy protons is available.  A full costing of the scheme and a detailed analysis of the required development work is beyond the scope of this manuscript.

The AWAKE experiment at CERN provides a valuable experimental testing ground for many aspects of the scheme that are most challenging, including the electron bunch injection into the modulated proton bunch, the subsequent acceleration to high energies while controlling the emittance, and the development of plasma sources with the required uniformity over the needed distance scales. Other aspects of the scheme, such as polarization preservation during plasma acceleration, would need to be demonstrated elsewhere.
This study serves as a reference for future potential applications of proton-driven wakefield acceleration. Developing an injector based on proton-driven plasma wakefields such as the one we describe would have the exciting additional benefit of demonstrating a new technology.  This could become important for future nuclear and particle-physics applications.

\section*{Acknowledgements}
The authors gratefully acknowledge the Gauss Centre for Supercomputing e.V. (www.gauss-centre.eu) for funding this project by providing computing time through the John von Neumann Institute for Computing (NIC) on the GCS Supercomputer JUWELS at Jülich Supercomputing Centre (JSC).

\section*{Declarations}
Data availability: simulation data is available upon reasonable request.

\bibliography{alive}

\end{document}